\documentclass[prb,twocolumn,showpacs]{revtex4}     
\usepackage{epsfig,amsmath} 

\begin{document}

\title{Spin-Polarized-Current State of Electrons in Bilayer Graphene} 

\author{Xin-Zhong Yan$^{1}$ and C. S. Ting$^2$}
\affiliation{$^{1}$Institute of Physics, Chinese Academy of Sciences, P.O. Box 603, 
Beijing 100190, China\\
$^{2}$Texas Center for Superconductivity, University of Houston, Houston, Texas 77204, USA}
 
\date{\today}

\begin{abstract}
We propose a model of spin-polarized-current state for electrons in bilayer graphene. The model resolves the puzzles as revealed by experiments that (a) the energy gap $E_{\rm gap}$ of the insulating ground state at the charge neutrality point (CNP) can be closed by a perpendicular electric field of either polarity, (b) $E_{\rm gap}$ increases significantly with increasing the magnetic field $B$, (c) the particle-hole spectrum is asymmetric in the presence of $B$, (d) there is a peak structure in the electric conductivity at small $B$ at the CNP, and (e) there are quantum Hall states stemming from lifting of degeneracy in the lowest Landau level. The model predicts that the ground state of the system close to the CNP is a ferrimagnet at finite $B$ and the Hall current is spin polarized.    
\end{abstract}


\maketitle

Recently, the bilayer graphene (BLG) has been studied extensively because of its potential application to new electronic devices.\cite{Ohta,Oostinga,McCann,Castro} Many experiments \cite{Weitz,Freitag,Velasco,Bao} performed on high quality suspended BLG samples have shown strong evidence for the gapped ground state of electrons at the charge neutrality point (CNP). The main experimental findings are: (i) the ground state is insulating with a gap that can be closed by a perpendicular electric field of either polarity, (ii) the gap grows with increasing magnetic field $B$ as $E_{\rm gap} = \Delta_0+\sqrt{a^2B^2 + \Delta_0^2}$ with $\Delta_0 \approx$ 1 meV and $a \approx$ 5.5 meVT$^{-1}$, (iii) the state is particle-hole asymmetric in the presence of the magnetic field $B$, (iv) there is a peak structure in the electric conductivity at small $B \approx $ 0.04T at the CNP, and (v) there are quantum Hall states ($\nu$ = 0, $\pm 1$, $\pm 2$ and $\pm 3$) stemming from lifting of degeneracy in the lowest Landau level. These experimental observations are still puzzles to the existing theories \cite{Min,Nandkishore,Jung,Zhang2,Gorbar,Nandkishore1,Zhang1,Zhu,Yan1,Yan2} including the models of the ferroelectric-layer asymmetric state \cite{Min,Nandkishore} or quantum valley Hall state (QVH),\cite{Zhang2} layer-polarized antiferromagnetic state (AF),\cite{Gorbar} quantum anomalous Hall state (QAH),\cite{Jung,Nandkishore1,Zhang1} quantum spin Hall state (QSH),\cite{Jung,Zhang1} and ordered-current state (OCS).\cite{Zhu,Yan1,Yan2} The QVH, QAH and QSH states all have been ruled out by the experiment.\cite{Velasco} It is shown that the AF state is not able to reproduce the gap growth with $B$.\cite{Yan2} The carrier density position of the gap given by the OCS deviates from the CNP at finite $B$ and the OCS cannot correctly explain (v). 

In this work, we propose a model of spin-polarized-current state (SPCS) for the electrons in BLG. We study the order parameters, the gap behavior, and the energy levels of the SPCS in the presence of the magnetic field. We will show that the experimental observations (i)-(v) stated above can be explained by the present theory. With the theory, we will also give new predictions.

{\it The Hamiltonian.} The unit cell of the BLG lattice shown in Fig. 1 contains atoms $a$ and $b$ on top layer, and $a'$ and $b'$ on bottom layer with lattice constant $a \approx 2.4$ \AA~ and interlayer distance $d \approx 3.34$ \AA. The energy of intralayer nearest-neighbor (NN) [between $a$ ($a'$) and $b$ ($b'$)] and interlayer NN (between $b$ and $a'$) electron hopping are $t$ and $t_1$, respectively. From the density-functional calculation \cite{Tatar} and the experiments,\cite{LMZ} the values of these quantities are determined in the ranges: 2.8 eV $< t <$ 3.1 eV and 0.27 eV $< t_1 <$ 0.4 eV. We here take $t$ = 3 eV and $t_1$ = 0.273 eV. The Hamiltonian of the continuum model for the noninteracting electrons is
\begin{eqnarray}
H_0&=&\sum_{vk\sigma}C^{\dagger}_{vk\sigma}H^0_{vk}C_{vk\sigma} \label {hm}
\end{eqnarray}
with $C^{\dagger}_{vk\sigma}=(c^{\dagger}_{avk\sigma},c^{\dagger}_{bvk\sigma}
,c^{\dagger}_{a'vk\sigma},c^{\dagger}_{b'vk\sigma})$ and $H^0_{vk}= \epsilon_0(s_vk_x\sigma_x-k_y\sigma_y)\tau_0-t_1(\sigma^-\tau^++\sigma^+\tau^-)$,
where $c^{\dagger}_{lvk\sigma}$ creates a spin-$\sigma$ electron of momentum $k$ in valley $v$ [= $K \equiv (4\pi/3a,0)$ or $K'=-K$] of sublattice $l$, $k$ is measured from the Dirac point $K$ ($K'$) and confined to a circle $k \leq 1/a$ in $K$ ($K'$) valley, $s_v = 1$ (-1) for $v = K$ ($K'$), $\epsilon_0 = \sqrt{3}t/2$, the Pauli matrices $\sigma$'s operate in $(a,b)$ or $(a',b')$ space, and $\tau$'s in the space of (top, bottom) layers. We hereafter use the units of $\epsilon_0 = a$ = 1. 

The interaction part of the Hamiltonian is
\begin{eqnarray}
H' = U\sum_{lj}\delta n_{lj\uparrow}\delta n_{lj\downarrow}+\frac{1}{2}\sum_{li\ne l'j}v_{li,l'j}\delta n_{li}\delta n_{l'j} \label {ith}
\end{eqnarray}  
where $\delta n_{li\sigma}=n_{li\sigma}-n/2$ is the number deviation of electrons with spin $\sigma$ from the average occupation $n/2$ at site $i$ of sublattice $l$, $\delta n_{li}=\delta n_{li\uparrow}+\delta n_{li\downarrow}$, and $U$ and $v$'s are the interactions between electrons. The off-site interactions here are given as
$v(r) = e^2[1-\exp(-q_0r)]/r$
where $r = |\vec r|$ with $\vec r$ as a vector from $li$ to $lj$, and $q_0$ is a parameter that approximately takes into account the wavefunction spreading effect in short range. According to the many-particle theory, since the exchange self-energy of electrons contains the screening due to the electronic charge fluctuations, we adopt the effective exchange interaction
\begin{eqnarray}
v_{xc}(r) = \frac{e^2}{r(1+\alpha q_sr+q^2_s r^2)} \label{int}
\end{eqnarray}  
where $q_s = 2\pi e^2\chi_0$ is the screening constant with $\chi_0 = t_1\ln 4/\pi(a\epsilon_0)^2$ the polarizability by the random-phase-approximation (RPA),\cite{Hwang} and $\alpha$ is an adjustable parameter. Note that the form of $v_{xc}(r)$ is consistent with the RPA in the limit $r \to \infty$. The total Hamiltonian $H_0 + H'$ satisfies the particle-hole symmetry.\cite{Yan}

{\it Self-energy of electrons.} The self-energy $\Sigma^{\sigma}_{ll'}(k)$ contains the Hartree and exchange terms. The off-diagonal part of the self-energy comes from the exchanges and results in a renormalization of $H^0_{vk}$. We will drop this part by supposing that it has already been included in $H^0_{vk}$. The Hartree terms in the diagonal part stem from the density orderings $\langle\delta n_{lj\sigma}\rangle$'s. In terms of the orderings of spin $m_l = (\langle\delta n_{lj\uparrow}\rangle-\langle\delta n_{lj\downarrow}\rangle)/2$ and charge $\rho_l = \langle\delta n_{lj\uparrow}\rangle+\langle\delta n_{lj\downarrow}\rangle$, we have $\langle\delta n_{lj\sigma}\rangle = \sigma m_l + \rho_l/2$ with $\sigma$ = + (-) for spin up (down). Since the charge ordering $\rho_l$ is the deviation from the average electron concentration $n$, those $\rho_l$'s satisfy the relations $\rho_a = -\rho_{b'}$ and $\rho_b = -\rho_{a'}$. The exchange self-energy in the diagonal part is due to the average $\langle c_{li\sigma} c^{\dagger}_{lj\sigma}\rangle = R_{l\sigma}(r)+iI_{l\sigma}(\vec r)$. The imaginary part $I_{l\sigma}(\vec r)$ is proportional to a current that breaks the time-reversal symmetry. In the previous work,\cite{Yan2} we neglected the spin dependence in $\langle c_{li\sigma} c^{\dagger}_{lj\sigma}\rangle$. Here, we keep the spin dependence in this average. Under the mean-field approximation and neglecting the terms of orders $\leq O(k)$, the self-energy in the diagonal part is obtained as
\begin{eqnarray}
\Sigma^{\sigma v}_{ll} = \epsilon_l-\sigma u_0m_l-s_v\Delta_{l\sigma}-v_c\delta/2 \label {slfe}
\end{eqnarray}  
where $\epsilon_l$ is due to the charge ordering, $\Delta_{l\sigma}$ stems from the current ordering, $\delta = n-1$, and $u_0$ and $v_c$ are effective interactions [see the supplementary material (SM)\cite{SM}]. In terms of $\rho_l$, $\epsilon_l$'s are given by $\epsilon_a = v_{aa}\rho_a + v_{ab}\rho_b$, $\epsilon_b = v_{bb}\rho_b + v_{ba}\rho_a$, $\epsilon_{b'}= -\epsilon_a$, and $\epsilon_{a'}= -\epsilon_b$. The interactions $v_{aa}$, $v_{bb}$ and $v_{ab} = v_{ba}$ are defined in SM.\cite{SM} The order parameters $\rho_l$, $m_l$ and $\Delta_{l\sigma}$ are determined by
\begin{eqnarray}
\rho_l &=& \frac{1}{2N}\sum_{v k\sigma}(\langle c^{\dagger}_{lvk\sigma}c_{lvk\sigma}\rangle-\langle c^{\dagger}_{\bar lvk\sigma}c_{\bar lvk\sigma}\rangle)                  \label {rho}\\
m_l &=& \frac{1}{2N}\sum_{v k\sigma}\sigma\langle c^{\dagger}_{lvk\sigma}c_{lvk\sigma}\rangle                  \label {ml}\\
\Delta_{l\sigma} &=& \frac{v_s}{N}\sum_{v k}s_v\langle c^{\dagger}_{lvk\sigma}c_{lvk\sigma}\rangle                  \label {dlt}
\end{eqnarray}  
where $N$ is the total number of the unit cells, the $k$ summations run over a single valley, the sublattice $\bar l$ means that $\bar a = b'$ and $\bar b = a'$ and vis-\`a-vis, and $v_s$ is an effective interaction (see SM\cite{SM}).

\begin{figure}
\vskip 8mm 
\centerline{\epsfig{file=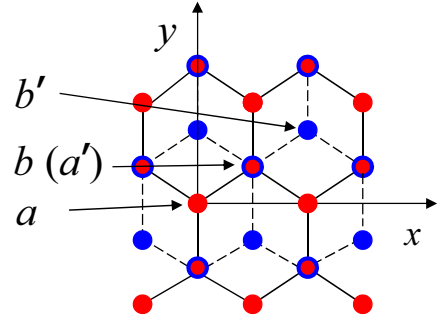,width=6. cm}}
\caption{(Color online) Top view of the bilayer graphene. Atoms $a$ ($a'$) and $b$ ($b'$) are on the top (bottom) layer.} 
\end{figure} 

{\it The SPCS at the CNP with B = 0.} At the CNP and in the absence of external electric and magnetic field, we expect the gap stems only from the current ordering and impose the conditions $\Delta_{l\sigma}=-\Delta_{\bar l\sigma}$ and $\Delta_{l\uparrow}=-\Delta_{l\downarrow}$ on the solution. The gap between the valence and conduction bands is $2|\Delta_{a\sigma}|$. To reproduce the experimental data $\Delta_0$ = 1 meV, $v_s$ needs to be 6.372. With this condition, the adjustable parameter $\alpha$ in $v_{xc}(r)$ given by Eq. (\ref{int}) is determined as 4.69. The other interaction parameters are determined as $u_0 = 6.38$, $v_c = 5.38$, $v_{aa} \approx v_{bb} = 3.3$ and $v_{ab} = v_{ba} = 6.58$ (by taking $q_0 = 0.5/a$, see SM\cite{SM}). With these parameters, we obtain $\rho_l = m_l$ = 0 except $\Delta_{l\sigma}$ being finite.
 
The relation $\Delta_{l\uparrow}=-\Delta_{l\downarrow}$ means that the current flows in opposite direction for opposite spin. Therefore, the system is in the spin-polarized-current state.
 
{\it The SPCS at finite B.} Under the magnetic field $B$ applied perpendicularly to the BLG plane, the vector potential is $\vec A = (0,Bx)$. By using the raising and lowering operators $a^{\dagger}$ and $a$ for the variable $x+k_y/B=(a^{\dagger}+a)/\sqrt{2B}$ and the operator $k_x = -i\nabla_x = i\sqrt{B/2}(a^{\dagger}-a)$, the operator $s_vk_x+i(k_y+Bx)$ in $H^0_{vk}$ becomes $i\sqrt{2B}a^{\dagger}$ for $v = K$ or $i\sqrt{2B}a$ for $v = K'$. The eigenfunction is expressed as $\psi^{\mu}_{vn\sigma} = \Phi_{vn}X_{vn\sigma}^{\mu}$ with $\mu$ as the band index, and $\Phi_{vn}$ (a $4\times 4$ diagonal matrix) and $X_{vn\sigma}^{\mu}$ (a 4-component vector normalized to unity) are defined as
\begin{eqnarray}
\Phi_{vn} &=& {\rm Diag}(i\phi_{n-1+s_v},
\phi_{n-1},
\phi_{n-1},
-i\phi_{n-1-s_v})\nonumber\\
X^{\mu}_{vn\sigma} &=& (x_{vn\sigma}^{1\mu},
x_{vn\sigma}^{2\mu},
x_{vn\sigma}^{3\mu},
x_{vn\sigma}^{4\mu})^t, \nonumber
\end{eqnarray}
where $\phi_n$ is the $n$th level wave function of a harmonic oscillator centered at $x_c=-k_y/B$, and the superscript $t$ means the transpose of the vector. Here, when the subscript $n$ of $\phi_n$ is negative, the corresponding component in $X_{vn\sigma}^{\mu}$ is understood as zero. Especially, for $n$ = 0, there is only one state of energy $\Sigma^{\sigma K}_{aa}$ ($\Sigma^{\sigma K'}_{b'b'}$) at $K$ ($K'$) valley with the electrons staying on $a$ ($b'$) sublattice. The vector $X^{\mu}_{vn\sigma}$ and the eigenenergy $E^{\mu}_{vn\sigma}$ are determined by
$H_{vn\sigma}X^{\mu}_{vn\sigma} = E^{\mu}_{vn\sigma}X^{\mu}_{vn\sigma}$,
with
\begin{eqnarray}
H_{vn\sigma}=\begin{pmatrix}
\Sigma^{\sigma v}_{aa}& \epsilon^+_{vn}&0&0\\
\epsilon^+_{vn}&\Sigma^{\sigma v}_{bb}&-t_1&0\\
0&-t_1&\Sigma^{\sigma v}_{a'a'}&\epsilon^-_{vn}\\
0&0&\epsilon^-_{vn}&\Sigma^{\sigma v}_{b'b'}\\
\end{pmatrix}\nonumber
\end{eqnarray}
and $\epsilon^{\pm}_{vn}=\sqrt{B(2n-1\pm s_v)}$. The $k$ summations in Eqs. (\ref{rho})-(\ref{dlt}) for self-consistently determining the order parameters are now changed to summations over the $y$-component momentum $k_y$ and the Landau states.\cite{Yan2}

\begin{figure} 
\centerline{\epsfig{file=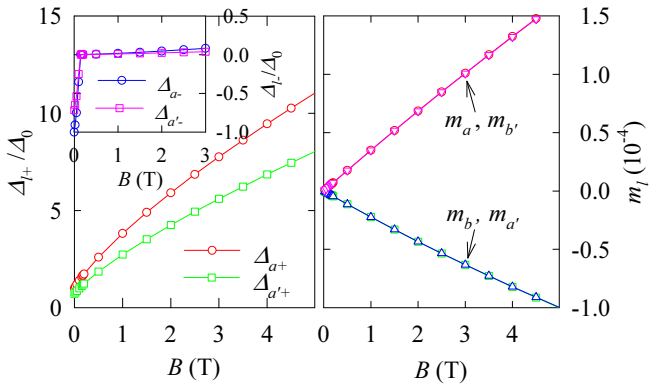,width=8.7 cm}}
\caption{(Color online) (Left) Order parameter $\Delta_{l\sigma}$ as function of magnetic field $B$. The main panel shows the result for spin-up electrons. The inset is for spin-down electrons. (Right) Spin polarization $m_l$ of sublattice $l$ as function of magnetic field $B$.} 
\end{figure} 

The solution at $\delta = 0$ to the order parameters $\Delta_{l\sigma}$ and $m_l$ are plotted in Fig. 2. At the CNP, these parameters satisfy the relationships: $\Delta_{l\sigma} = -\Delta_{\bar l\sigma}$ and $m_l = m_{\bar l}$, while the charge ordering parameters $\rho_l$ vanish. As shown in left panel of Fig. 2, the magnitudes of the current parameters $\Delta_{l+} = \Delta_{l\uparrow}$ for spin-up electrons increases with $B$, but the magnitude of $\Delta_{l-} = \Delta_{l\downarrow}$ for spin-down electrons decreases with $B$ in a small interval of $B$ close to zero; $\Delta_{l-}$ vanishes at $B \approx 0.15$ T and then very slowly increases with $B$. The behaviors of $\Delta_{l\sigma}$ can be understood by simply looking into the property of the $n$ = 0 state. As noted above, the energy of the state is $\Sigma^{\sigma K}_{aa} = -\sigma m_au_0-\Delta_{a\sigma}$ at $K$ valley or $\Sigma^{\sigma K'}_{b'b'} = -\sigma m_{b'}u_0+\Delta_{b'\sigma}$ at $K'$ valley. The energy is negative for spin-up electrons, while it is positive for spin-down electrons. At zero temperature and at the CNP, the latter state is empty. Therefore, the magnetic field enhances $\Delta_{l\uparrow}$ but suppresses $\Delta_{l\downarrow}$.  

Because there are more negative energy states for spin-up electrons than for spin-down electrons at finite $B$ and at the CNP, the system has a total net spin. It is seen from right panel of Fig. 2, the system is a ferrimagnet with the sublattices $a$ and $b'$ being equally spin-up ordered and the $b$ and $a'$ sublattices spin-down ordered. The magnitude of the spin polarization $m_l$ is approximately linear in $B$. The magnetization comes solely from the orbital current ordering but not the Zeeman splitting. The Zeeman splitting has been neglected here because the orbital effect is about 46 times larger than it. The `spin-up' here merely means its current ordering parameter $\Delta_{a\uparrow}$ is positive.  

\begin{figure} 
\centerline{\epsfig{file=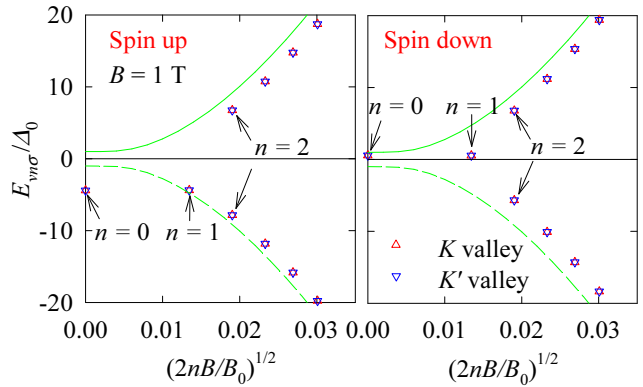,width=8.8cm}}
\caption{(Color online) Landau levels $E^{\mu}_{vn\sigma}$ in the valence and conduction bands at $B$ = 1 T. The lines represent the continuum conduction (solid) and valence (dashed) bands at $B = 0$ with momentum $k$ as the abscissa.} 
\end{figure} 

The Landau levels in the conduction and valence bands close to zero at $B$ = 1 T are shown in Fig. 3. At the CNP, because of $\rho_l$ = 0 and $m_l = m_{\bar l}$ and $\Delta_{l\sigma} = -\Delta_{\bar l\sigma}$, the energy levels are degenerated for exchanging the two valleys. On the other hand, the levels are different for different spins because of the spin polarization and the different current orderings. The obvious difference appears at the levels of $n$ = 0 and 1. The energy levels of $n$ = 1 are determined by the upper-left (lower-right) $3\times 3$ matrix of $H_{K1\sigma}$ ($H_{K'1\sigma}$). To order $O(B)$, the level of $n$ = 1 nearly degenerated with the level $\Sigma^{\sigma K}_{aa}$ of $n$ = 0 is obtained approximately as
\begin{eqnarray}
E_{K1\sigma}\approx\Sigma^{\sigma K}_{aa}+2B(\Sigma^{\sigma K}_{a'a'}-\Sigma^{\sigma K}_{aa})/t^2_1.  \label {ln1} 
\end{eqnarray}
The perturbation $E_{K1\sigma}-\Sigma^{\sigma K}_{aa}$ is positive for spin-up electrons but negative for spin-down electrons. By viewing the energy levels, the energy gap at the CNP is found as the difference between $E^c_{K1\downarrow}$ in the conduction band and $E^v_{K2\downarrow}$ in the valence band,
\begin{eqnarray}
E_{gap}=E^c_{K1\downarrow}-E^v_{K2\downarrow}.   \label{egap}
\end{eqnarray}
 
{\it Comparison with experiments.} (i) By experiment,\cite{Velasco} the gap is measured through the electric conductivity with a source-drain voltage applied to the sample. During such an electric transport process, the spin should not be altered and the gap should be given by Eq. (\ref{egap}). The gap is shown as a function of $B$ in Fig. 4. Except a dip at $B \approx 0.15$ T, the theory reproduces satisfactorily the experimental result.\cite{Velasco} (ii) Though the dip is not observed in Ref. \onlinecite{Velasco}, the appearance of the dip is in qualitatively agreement with the observation by Weitz {\it et al.}.\cite{Weitz} The latter experiment shows that there is peak structure in the electric conductivity at $|B| \approx 0.04$ T, which implies the dip in the energy gap. (iii) On the other hand, the energy bands have no particle-hole symmetry, which is in agreement with the experiment.\cite{Velasco} (iv) Because the levels of $n$ = 0 and 1 of spin-up electrons in the valence band are occupied while their counterparts of spin-down electrons in the conduction band are empty, we have obtained the insulating state with $\nu$ = 0 at the CNP. This is different from the previous OCS model \cite{Yan2} by which the levels $n$ = 0 and 1 are degenerated for both spins and all are occupied (empty) when they are negative (positive). Thus, the electron density of the gapped state given by the previous model cannot not be viewed at the CNP. (v) Finally, the gap can be closed by perpendicular electric field in either direction. To see it, we apply voltages $\pm V$ respectively to the top and bottom layers. This causes charge polarization between the two layers. The quantity $\epsilon_l$ in Eq. (\ref{slfe}) now includes the voltage and the charge ordering effect (and is finite). For positive (negative) $V$, we get positive (negative) $\epsilon_a$. The level $\Sigma^{\uparrow K}_{aa}$ ($\Sigma^{\uparrow K'}_{b'b'}$) in the valence band raises, while the level $\Sigma^{\downarrow K'}_{b'b'}$ ($\Sigma^{\downarrow K}_{aa}$) in the conduction band decreases. At certain $V$, the phase transition with the particle distribution changing in the top level of the valence band and bottom level of conduction band happens and the gap closes.\cite{Yan2} 

\begin{figure} 
\centerline{\epsfig{file=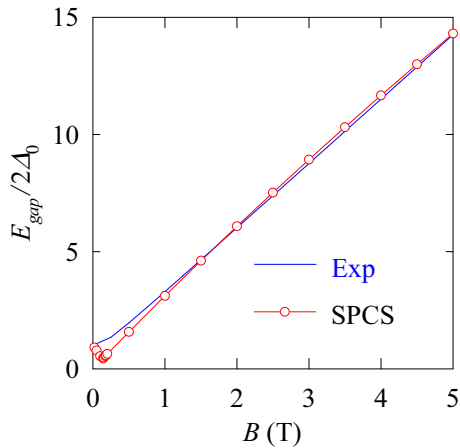,width=6.5cm}}
\caption{(Color online) Energy gap $E_{gap}$ at CNP as function of $B$ compared with the experimental result (Exp) \cite{Velasco}.} 
\end{figure} 

{\it Quantum Hall states (QHS) of integer $|\nu| \leq 4$.} By doping electrons, the level $E^c_{K1\downarrow}$ or $E^c_{K'1\downarrow}$ is firstly filled with spin-down electrons. The occupation of level $E^c_{K1\downarrow}$ ($E^c_{K'1\downarrow}$) close to $\Sigma^{\downarrow K}_{aa}$ ($\Sigma^{\downarrow K'}_{b'b'}$) means that the sublattice $a$ ($b'$) is mostly occupied. Therefore, if the level $E^c_{K1\downarrow}$ is filled, there will exist charge ordering with $\rho_a > 0$ and $\rho_b < 0$, resulting in $\epsilon_a < 0$ and $E^c_{K1\downarrow} < E^c_{K'1\downarrow}$. This is the state of $\nu$ = 1. Analogously, we can analyze the other states of integer $|\nu| \leq 4$. The key point is that under the carrier doping the valley degeneracy of the Landau levels is lifted by the charge orderings $\rho_l \ne 0$ (see SM\cite{SM}). The appearance of these QHS is in qualitatively agreement with the experimental observations.\cite{Elferen,Velasco1} 

{\it Prediction.} As stated above, the system is a ferrimagnet at the CNP under the magnetic field. Moreover, since the Hall states of $\nu$ = 1, 2, 3 and 4 correspond to the occupations of levels of $n$ = 0 and 1 in the conduction band with spin-down electrons, the Hall current in these states is spin-down polarized. On the other hand, the Hall current in the states of $\nu$ = -1, -2, or -3 is spin-up polarized because the states of $n$ = 0 and 1 in the valence band are for spin-up electrons.

{\it Summary.} On the basis of the four-band continuum Hamiltonian, we have proposed a model of spin-polarized-current state for the interacting electrons in BLG. The model can explain the experimental observations (i)-(v) as stated in the beginning of the paper. The model predicts that (a) the ground state of the system close to the CNP is a ferrimagnet at finite $B$ and (b) the Hall current is spin polarized.    

This work was supported by the National Basic Research 973 Program of China under Grants No. 2011CB932700 and No. 2012CB932302, and the Robert A. Welch Foundation under Grant No. E-1146.

%%%%%%%
\pagebreak

\widetext
\begin{center}
\textbf{Supplementary Material for `Spin-polarized-current state of electrons in bilayer graphene'} 

\end{center}

\setcounter{page}{1}
\setcounter{equation}{0}
\setcounter{figure}{0}

{\it 1. Approximation for the self-energy.} The interaction part of the Hamiltonian is
\begin{eqnarray}
H' = U\sum_{lj}\delta n_{lj\uparrow}\delta n_{lj\downarrow}+\frac{1}{2}\sum_{li\ne l'j}v_{li,l'j}\delta n_{li}\delta n_{l'j} \label {ith}
\end{eqnarray}  
where $\delta n_{li\sigma}=n_{li\sigma}-n/2$ is the number deviation of electrons with spin $\sigma$ from the average occupation $n/2$ at site $i$ of sublattice $l$, $\delta n_{li}=\delta n_{li\uparrow}+\delta n_{li\downarrow}$, $U$ is the on-site interaction, and $v_{li,l'j}$ is the interaction between electrons at sites $li$ and $l'j$. 

According to the renormalized-ring-diagram approximation \cite{Yan1}, the self-energy of electrons is given in Fig. 1. Here, for simplifying our calculation, we use an effective static exchange interaction as given by Eq. (3) in the main text taking into account the screening effect due to electronic charge fluctuations.

The Hartree terms in the diagonal part of the self-energy stem from the density orderings $\langle\delta n_{lj\sigma}\rangle$'s. In terms of the orderings of spin $m_l = (\langle\delta n_{lj\uparrow}\rangle-\langle\delta n_{lj\downarrow}\rangle)/2$ and charge $\rho_l = \langle\delta n_{lj\uparrow}\rangle+\langle\delta n_{lj\downarrow}\rangle$, we have $\langle\delta n_{lj\sigma}\rangle = \sigma m_l + \rho_l/2$ with $\sigma$ = + (-) for spin up (down). Since the charge ordering $\rho_l$ is the deviation from the average electron concentration $n$, those $\rho_l$'s satisfy the relations $\rho_a = -\rho_{b'}$ and $\rho_b = -\rho_{a'}$. The Hartree approximation reads
\begin{eqnarray}
H' &\approx& \sum_{li\sigma}\left(U\rho_l/2-\sigma Um_l+\sum_{l'j\ne li}v_{li,l'j}\rho_{l'}\right)\delta n_{li\sigma}. \nonumber
\end{eqnarray}  
The Hartree term in the self-energy is 
\begin{eqnarray}
\Sigma^{\sigma H}_{ll} &=& U\rho_l/2-\sigma Um_l+\sum_{l'j\ne li}v_{li,l'j}\rho_{l'} \nonumber\\
 &=& (u_{ll}+U/2)\rho_l+u_{l\tilde l}\rho_{\tilde l}-\sigma Um_l \label {ht}
\end{eqnarray}
where $\tilde l$ means that $\tilde a = b$, $\tilde b = a$, $\tilde a' = b'$, and $\tilde b' = a'$, and $u_{ll}$ and $u_{l\tilde l}$ are interaction parameters. The parameters $u_{aa}$ and $u_{ab} = u_{ba}$ are defined as
\begin{eqnarray}
u_{aa} &=& -v(|\vec r_0|)+\sum_{\vec r\ne 0}[v(r)-v(|\vec r +\vec r_0|)] \nonumber\\
u_{ab} &=& \sum_{\vec r}[v(|\vec r+\vec r_1|)-v(|\vec r +\vec r_2|)] \nonumber
\end{eqnarray} 
where the $\vec r$-summation runs over the $a$ sublattice, and
$\vec r_0 = (1,1/\sqrt{3},d)$ and $\vec r_1 = (1/2,1/2\sqrt{3},0)$ and $\vec r_2 = (1/2,1/2\sqrt{3},d)$ are the vectors from atom $a$ to respectively atoms $b'$, $b$, and $a'$ in the unit cell. The parameter $u_{bb}$ is defined by 
\begin{eqnarray}
u_{bb} &=& -v(d)+\sum_{\vec r\ne 0}[v(r)-v(|\vec r +\vec d|)] \nonumber
\end{eqnarray}
where $\vec d = (0,0,d)$ with $d$ as the interlayer spacing, and the $\vec r$-summation runs over the $b$ sublattice.    

\begin{figure} 
\vskip 5mm
\centerline{\epsfig{file=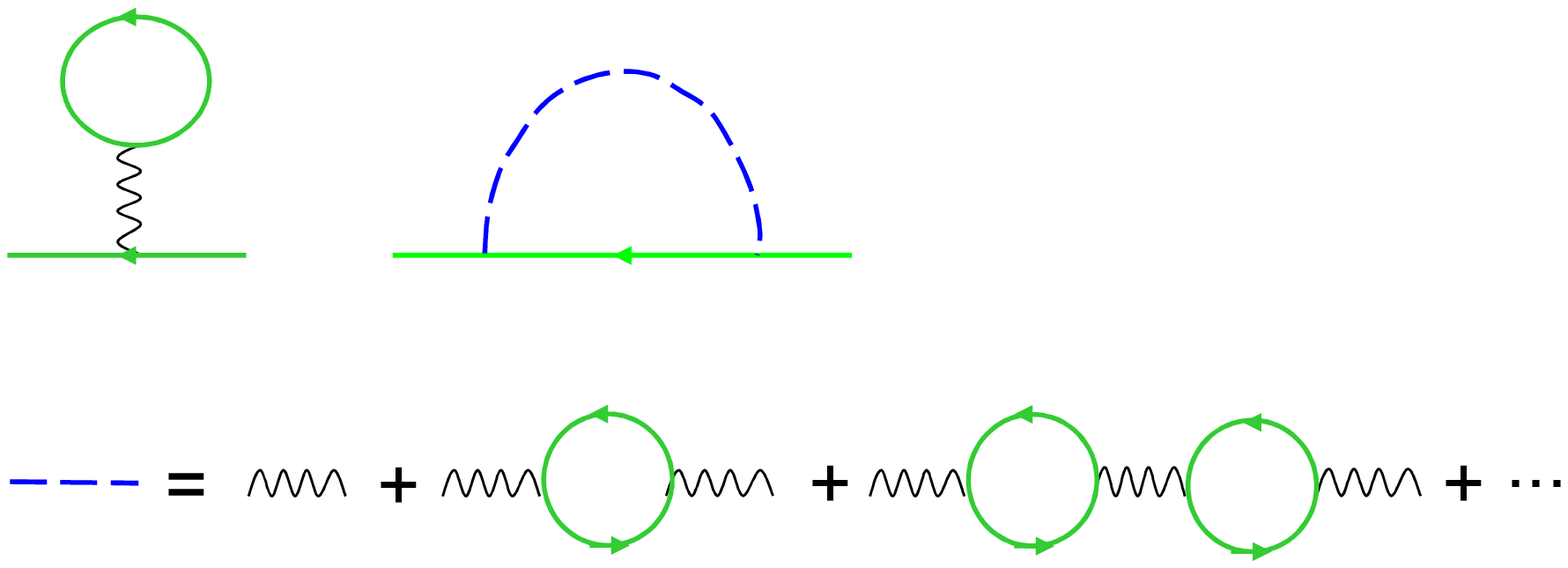,width=8.5cm}}
\caption{(Color online) Self-energy of electrons. The solid line with an arrow represents the Green's function of electrons. The waveline is the Coulomb interaction $v(r)$. The dashed line is the exchange interaction.} 
\end{figure} 

On the other hand, according to the treatment in the previous work \cite{Yan2}, the exchange self-energy in the diagonal part can be obtained as 
\begin{eqnarray}
\Sigma^{\sigma v,xc}_{ll} = -(\delta/2+\rho_l/2+\sigma m_l)v_c -s_v\Delta_{l\sigma}. \label {xc}
\end{eqnarray}
with   
\begin{eqnarray}
\Delta_{l\sigma} &=& \frac{v_s}{N}\sum_{v k}s_v\langle c^{\dagger}_{lvk\sigma}c_{lvk\sigma}\rangle                  \nonumber\\
v_c &=& \sum_{\vec r\ne 0}v_{xc}(r)\cos^2(\vec K\cdot\vec r) \nonumber\\
v_s &=& \sum_{\vec r\ne 0}v_{xc}(r)\sin^2(\vec K\cdot\vec r) \nonumber
\end{eqnarray} 
where $\delta$ is the concentration of doped electrons, the $\vec r$-summation runs over the $a$ sublattice, and $v_{xc}(r)$ is defined in the main text.

By summing $\Sigma^{\sigma H}_{ll}$ and $\Sigma^{\sigma v,xc}_{ll}$, we obtain 
\begin{eqnarray}
\Sigma^{\sigma v}_{ll} = \epsilon_l-\sigma u_0m_l-s_v\Delta_{l\sigma}-v_c\delta/2 \label {slfe}
\end{eqnarray}  
where $\epsilon_l$'s are given by $\epsilon_a = (u_{aa}+U/2-v_c/2)\rho_a + u_{ab}\rho_b$, $\epsilon_b = (u_{bb}+U/2-v_c/2)\rho_b + u_{ba}\rho_a$, $\epsilon_{b'}= -\epsilon_a$, and $\epsilon_{a'}= -\epsilon_b$, and $u_0 = U +v_c$. 

\begin{figure}[t] 
\centerline{\epsfig{file=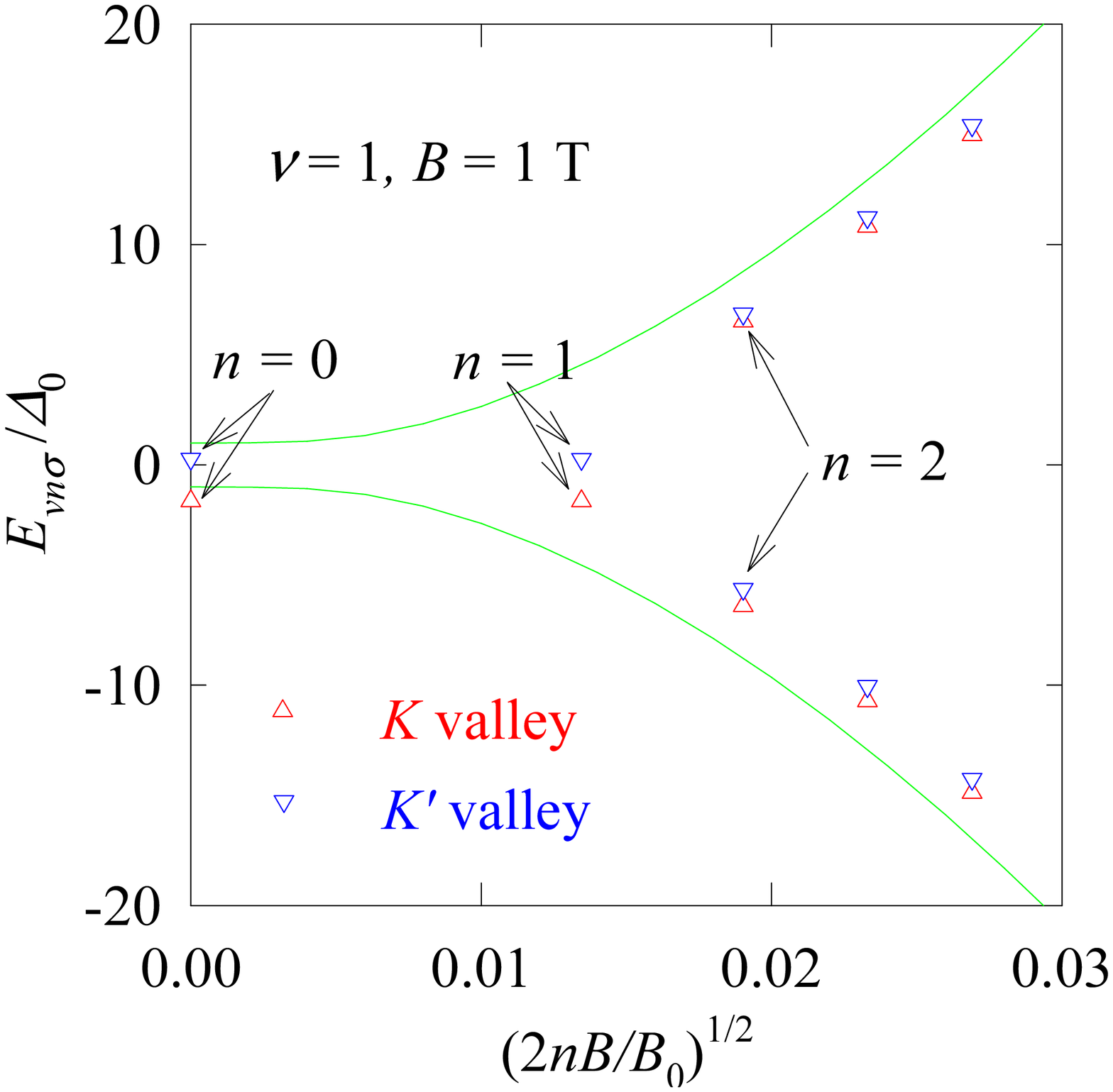,height=6.5cm}}
\caption{(Color online) Landau levels $E_{vn\sigma}$ for spin-down electrons $\sigma = \downarrow$ at the two valleys $v = K$ and $K'$ of quantum Hall state $\nu = 1$ at $B$ = 1 T.} 
\vskip 3mm
\end{figure} 

With the parameter $\alpha = 4.69$ in $v_{xc}(r)$, we obtain the interaction parameters $v_c = 5.38$, $v_s = 6.372$, reproducing the experimental gap $\Delta_0$ = 1 meV. Since a strong on-site interaction $U$ leads to the AF state and the state does not explain the experimental observation, the strength of $U$ should be weak. We here take $U = \epsilon_0 \approx 2.66 v(a)$ and obtain $u_0 = 6.38$. 

\begin{figure}[t]
\centerline{\epsfig{file=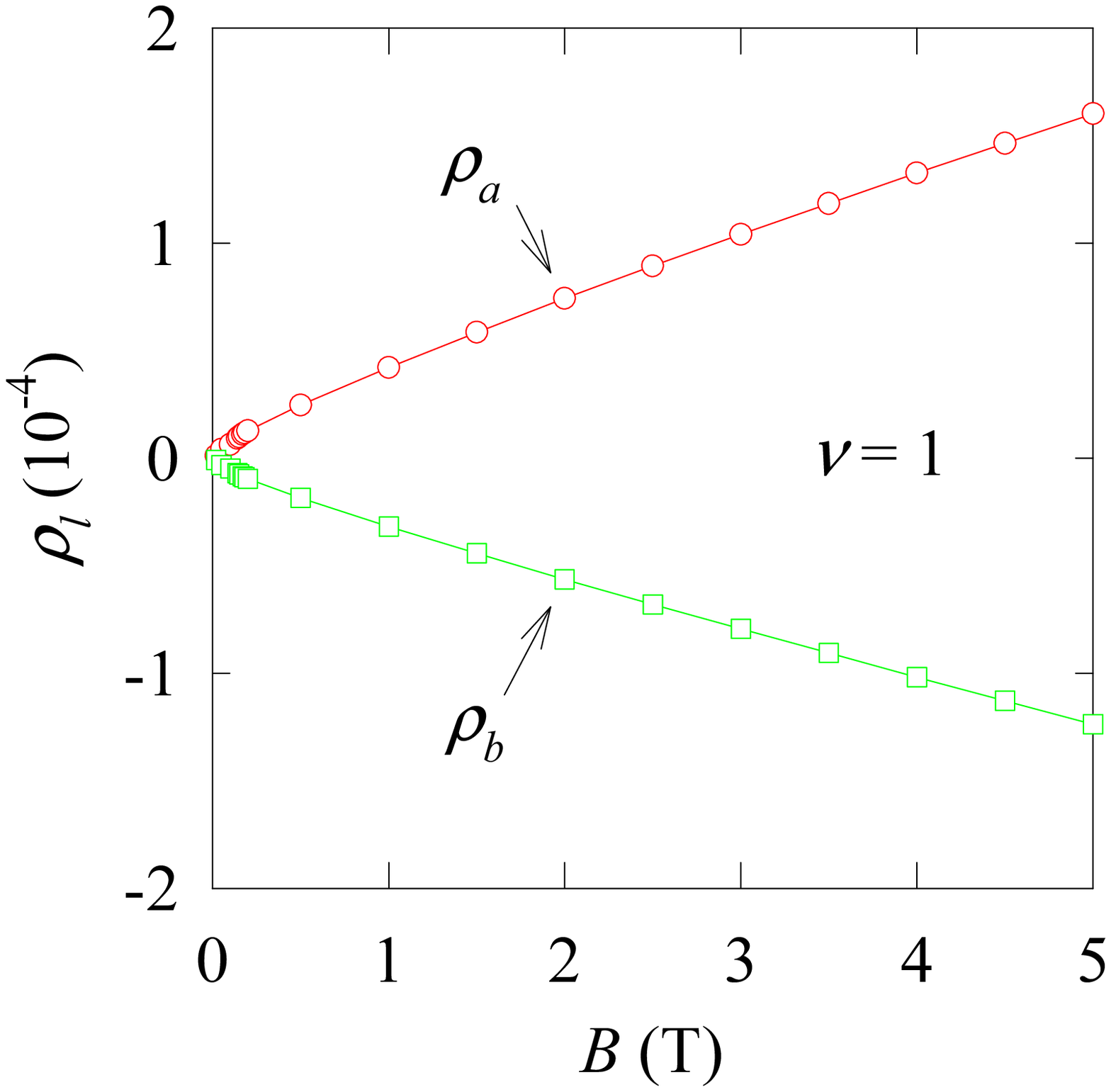,height=6.5cm}}
\caption{(Color online) Charge orderings $\rho_l$ as functions of magnetic field $B$ of the $\nu = 1$ quantum Hall state.} 
\end{figure} 

For $q_0 \le 0.58/a$, the interaction $v(r)$ does not result in charge orderings in the system at the charge neutrality point (CNP) (and the results at the CNP do not depend on $q_0$ for $q_0 \le 0.58/a$). We here take $q_0 = 0.5/a$ and get $u_{aa} \approx u_{bb} = 5.495$ and $u_{ab} = 6.577$ from $v(r)$. The interactions $v_{aa}=v_{bb}$ and $v_{ab}=v_{ba}$ in the text are given as
\begin{eqnarray}
v_{aa} &=& u_{aa}+U/2-v_c/2 = 3.3, \nonumber\\
v_{ab} &=& u_{ab} = 6.58. \nonumber
\end{eqnarray} 

{\it 2. Landau levels of the $\nu = 1$ QHS.} By doping electrons, the Landau level $E^c_{K1\downarrow}$ or $E^c_{K'1\downarrow}$ can be firstly filled with spin-down electrons. We here consider the case that the level $E^c_{K1\downarrow}$ is firstly filled. This is the case that the sublattice $a$ is mostly occupied. Figure 2 shows the Landau levels $E_{vn\sigma}$ of spin-down electrons $\sigma = \downarrow$ of the $\nu = 1$ quantum Hall state (QHS) at $B$ = 1 T. Due to the charge orderings, the valley ($v = K, K'$) degeneracy is lifted. Shown in Fig. 3 are the corresponding charge orderings $\rho_l$ as functions of the magnetic field $B$.

\end{document}